\documentclass[twocolumn,10pt]{article}

\usepackage{amsmath}
\usepackage{amssymb}

\usepackage{graphicx}

\usepackage{cite}
\usepackage{color} 

\usepackage[labelfont=bf,labelsep=period,justification=raggedright]{caption}

\makeatletter
\renewcommand{\@biblabel}[1]{\quad#1.}
\makeatother

\date{}

\begin{document}
\onecolumn
% Title must be 150 characters or less
\begin{flushleft}
{\Large
\textbf{Patterns of subnet usage reveal distinct scales of regulation\\[.1cm] in the transcriptional regulatory network of
  \textit{Escherichia coli}}}\\[.5cm]

% Insert Author names, affiliations and corresponding author email.
Carsten Marr$^{1,\ast}$, Fabian J.~Theis$^{1,2}$, Larry
S.~Liebovitch$^{3}$, Marc-Thorsten H\"utt$^{4}$
\\[.5cm]
\bf{1} Institute for Bioinformatics and Systems Biology, Helmholtz
Zentrum M\"unchen - German Research Center for Environmental Health, Neuherberg,
Germany
\\
\bf{2} Institute for Mathematical Sciences, Technische Universit\"at M\"unchen, Garching, Germany
\\
\bf{3} Center for Complex Systems and Brain Sciences, Florida Atlantic
University, Boca Raton, Florida, USA
\\
\bf{4} Computational Systems Biology, School of Engineering and
Science, Jacobs University Bremen, Bremen, Germany
\\
$\ast$ E-mail: carsten.marr@helmholtz-muenchen.de
\end{flushleft}

% Please keep the abstract between 250 and 300 words
\section*{Abstract}
The set of regulatory interactions between genes, mediated by
transcription factors, forms a species' transcriptional regulatory
network (TRN). By comparing this network with measured gene expression
data one can identify functional properties of the TRN and gain
general insight into transcriptional control.
We define the subnet of a node as the subgraph consisting of all nodes
topologically downstream of the node, including itself. Using a large
set of microarray expression data of the bacterium \textit{Escherichia
  coli}, we find that the gene expression in different subnets
exhibits a structured pattern in response to environmental changes and
genotypic mutation.  Subnets with less changes in their expression
pattern have a higher fraction of feed-forward loop motifs and a lower
fraction of small RNA targets within them.
Our study implies that the TRN consists of several scales of
regulatory organization: 1) subnets with more varying gene expression
controlled by both transcription factors and post-transcriptional RNA
regulation, and 2) subnets with less varying gene expression having
more feed-forward loops and less post-transcriptional RNA regulation.

% Please keep the Author Summary between 150 and 200 words
% Use first person. PLoS ONE authors please skip this step. 
% Author Summary not valid for PLoS ONE submissions.   
\section*{Author Summary}
Bacterial cells can adapt to various genomic mutations and
intriguingly many environmental changes. They do this by adjusting
their gene expression profile to meet the requirements of a new
condition. In this work, we study the interplay of different
mechanisms of gene regulatory control driving this adaptation in the
bacterium \textit{Escherichia coli}. We deconstruct the network of all
transcription factor mediated regulatory interactions into subnets,
topologically defined subgraphs which we expect to act as information
processing units. Indeed, we find that many subnets react coordinately
to cellular stress, and are used by the cells to account for
mutations. In these subnets, we also find many small RNA targets. 
In contrast, those subnets that do not act in a coordinated fashion
are highly enriched with feed-forward loops, a 3-node network motif with
important information processing properties. Our approach
reveals correlations and anti-correlations of three scales of
regulatory control: subnets, feed-forward loops, and small RNA.

\twocolumn
\section*{Introduction}
An interesting topological feature of the transcriptional regulatory
network (TRN) of the bacterium \textit{Escherichia coli} is its almost
tree-like structure with only few loops (see \cite{yu06_pnas} for a
  detailed discussion and comparison with the TRN of the yeast
  \textit{Saccharomyces cerevisiae}). This observation has several
consequences. First, hierarchical levels in the network can be
meaningfully defined and analyzed. Second, it leads to the question,
on which level of organization information processing takes place in
the TRN given a dominant directed flow dictated by the network's
architecture. On a local scale, substructures in the TRN that appear
significantly more often than in corresponding randomized
networks---so-called network motifs \cite{shen-orr02, milo02}---have
been found to match specific information processing
steps. Particularly feed-forward loops have been theoretically
proposed \cite{mangan03_pnas} and experimentally supported
\cite{mangan03_j-mol-biol,mangan06_j-mol-biol} to function as
noise-suppression units and delay devices.

\begin{figure}[!ht]
  \includegraphics[width=\columnwidth]{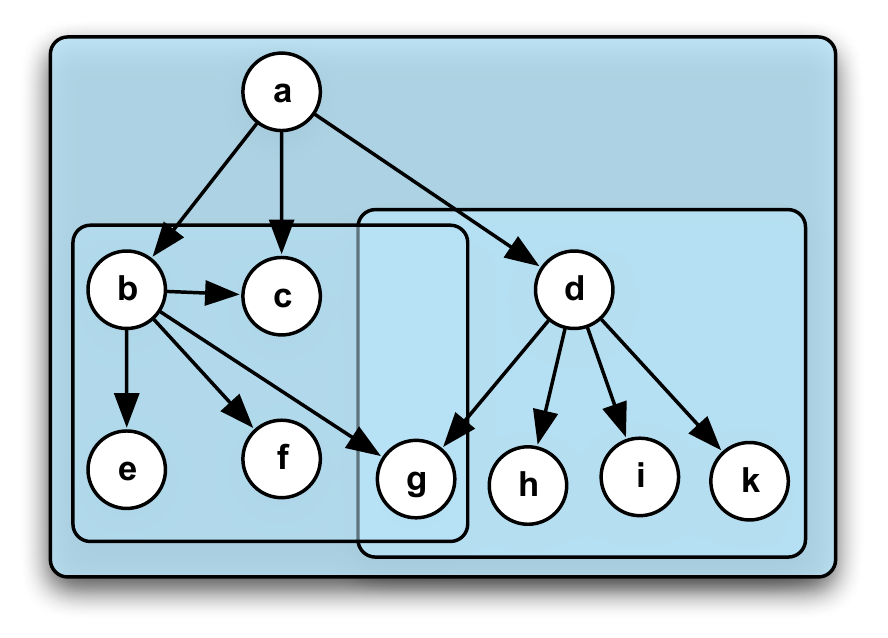}
  \caption{{\bf Illustration of the subnet approach.} A subnet is defined as
    the subgraph induced by all nodes downstream of a root node,
    including the root node. The network in this figure contains three
    subnets: the subnet of root \textbf{a} comprises all nodes in the
    network, organized in three hierarchical layers. The subnet of root
    \textbf{b} contains \textbf{b} and all downstream nodes
    \textbf{c}, \textbf{e}, \textbf{f}, \textbf{g}. The subnet of root
    \textbf{d} contains \textbf{d}, \textbf{g}, \textbf{h},
    \textbf{i}, \textbf{k}. Notably, the subnet of root \textbf{a}
    contains a feed-forward loop formed by nodes \textbf{a},
    \textbf{b}, and \textbf{c}, while the subnet of root \textbf{d}
    constitutes a single input motif.}
  \label{fig:fflvssubnet}
\end{figure}

Here we dissect the TRN into topological modules. We define the
subnet of a node (root) as the subgraph consisting of all nodes
topologically downstream of the root, including the root node itself
(see Figure \ref{fig:fflvssubnet} for an illustration of the
concept). Subnets can extend over multiple hierarchical layers if they
contain a hierarchy of transcriptions factors (TFs). Moreover, they
can overlap if genes are regulated by TFs from different subnets. Some
network motifs such as the feed-forward loop or the single input motif
are subnets themselves and therefore fully contained in at least one
subnet. This approach is possible due to the topological properties of
the \textit{E.~coli} TRN: apart from the few small cycles in the
network (see Results), most subnets are directed acyclic graphs. 

The search for the imprint of the transcriptional regulatory network
in gene expression profiles is a search for very weak signals, often
masked by the broad range of additional biological processes (beyond
the regulation via transcription factors) shaping the expression of a
gene. In two previous studies \cite{herrgard03,gutierrez03}, the
consistency between expression profiles and pairwise interactions in
the TRN has been shown to be surprisingly low. The consistency on a
larger scale has been studied for a specific type of subnets, named
`origons' \cite{balazsi05}. There, the authors find that genes in some
origons are selectively affected by specific environmental signals.
In this contribution, we study patterns of subnet usage for two
markedly different genome-wide gene expression data sets. As is
\cite{balazsi05}, we use microarray expression profiles from the ASAP
database, where wild-type expression under standard growth conditions
is compared to a variety of profiles with external stimuli and genetic
alterations.  As a second data set, we use the time-course data of
\cite{sangurdekar06_genome-biol}. Here, \textit{E.~coli} strains are
exposed to different media and stresses, and profiled at up to 16 time
points. We analyze subnets with respect to their responsiveness to
altered conditions in both data sets and classify them according to
the observed subnet usage patterns.

\textit{E.~coli} employs different scales of regulatory control to
establish homeostasis (see, e.g., \cite{blot06}) or to adapt to
external stimuli. Recently, we introduced the concept of digital and
analog control to differentiate between the regulatory response
coordinated by dedicated TFs and DNA architectural proteins,
respectively \cite{marr08_bmc-syst-biol}. We found that as soon as one
form is limited (by TF mutations or changes in the DNA superhelicity),
the other form of control compensates, exhibiting a balance of
regulatory control. An analysis employing methods from point process
statistics has been able to further support the interplay of digital
and analog control by analyzing gene distributions
\cite{sonnenschein09_bmc-syst-biol}. In the following, we want to
delineate the interplay between the subnet usage as a TF mediated,
topologically based form of control, and two other scales of
regulatory control: translational inhibition and mRNA degradation
induced by small non-coding RNAs (sRNAs) and the dynamic coordination
of nodes connected in a feed-forward loop.

\begin{figure*}[!ht]
\includegraphics[width=2\columnwidth]{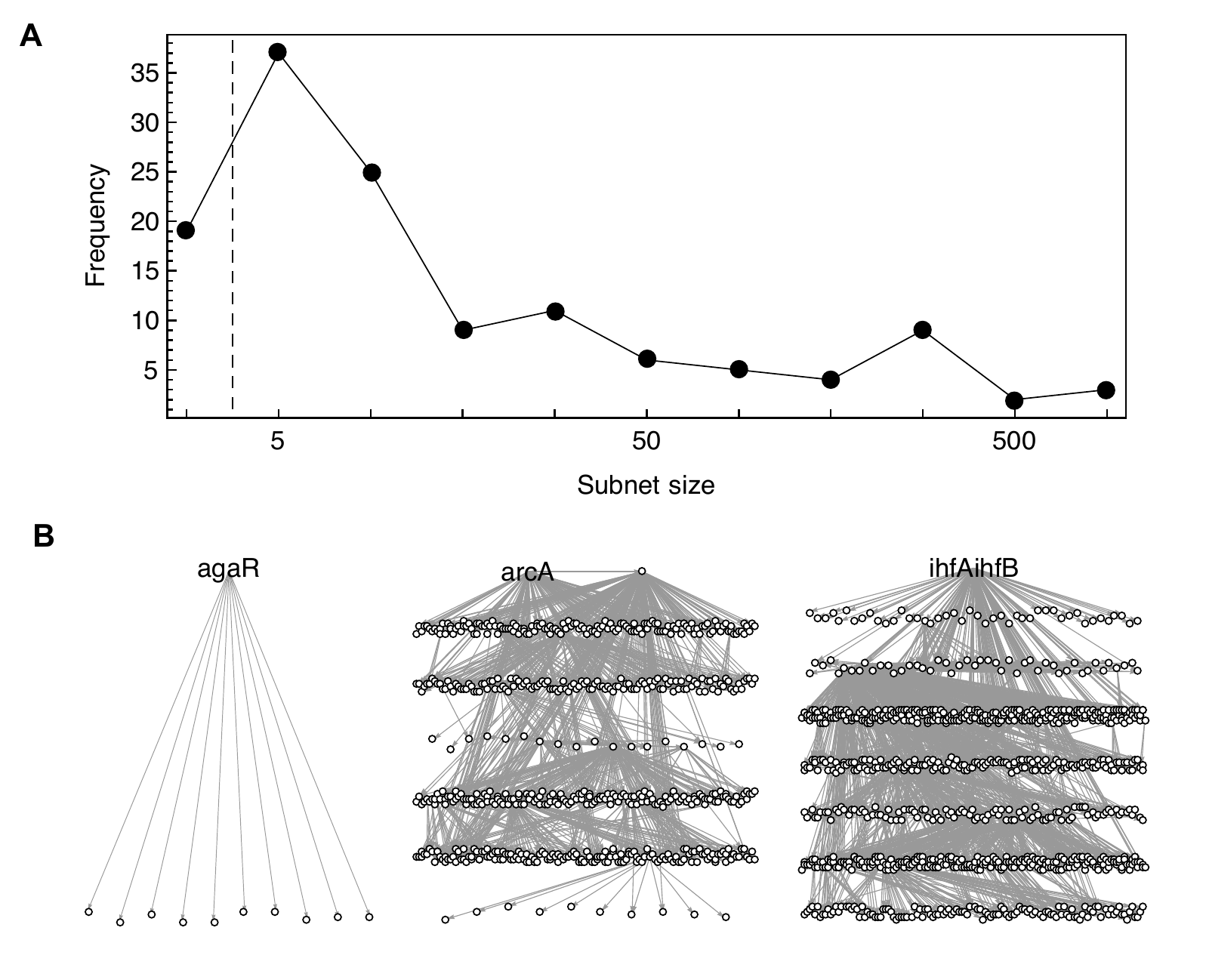}
\caption{{\bf Subnets in the transcriptional regulatory network (TRN)
    of \textit{E.~coli}.} The TRN can be decomposed into subnets,
  defined by the root node, comprising all nodes topologically
  downstream. (A) Histogram of subnet sizes, binned on a logarithmic
  scale. We only consider subnets of five nodes or more to allow for
  significantly enriched subnets. (B) The \textit{ihfAihfB} subnet is
  the largest subnet in the TRN, comprising 1021 nodes organized in
  seven hierarchical levels. The transcription factors \textit{arcA}
  and \textit{fnr} regulate each other and therefore share 650
  downstream nodes. The \textit{agaR} subnet has only 11 nodes,
  organized as a single input motif.}
  \label{fig:examples}
\end{figure*}

\section*{Results}
\subsection*{Networks}
We consider the most complete prokaryotic TRN available, the TRN of
the bacterium \textit{E.~coli}. Nodes in our network correspond to
genes (and the respective TF) while a directed edge represents a
regulatory interaction mediated by a TF.  Based on the version 6.3 of
the Regulon database \cite{gama-castro08_nucleic-acids-res}, the TRN
comprises 1515 nodes and 3171 links, with 162 regulators (i.e.~nodes
which regulate at least one other gene) and 1432 target nodes
(i.e.~nodes which are regulated by at least one other gene).

We dissect the TRN into subnets, defined as subgraphs consisting of a
root node with at least one regulatory interaction, and all downstream
nodes (see Figure \ref{fig:fflvssubnet}A for an illustrative example
network consisting of three subnets). The 162 subnets of the TRN are
overlapping and of very different sizes and hierarchical complexities
(see the frequency distribution of subnet sizes in Figure
\ref{fig:examples}A and the histogram of relative subnet overlap in
Figure 2 in Text S1). Let us consider three examples: the
\textit{ihfAihfB} subnet (see Figure \ref{fig:examples}B and Figure 1
in Text S1 for a highly resolved version) is the largest
subnet in the \textit{E.~coli} TRN with 1021 downstream nodes, among
them many regulators, organized in seven hierarchical levels. For the
genes \textit{ihfA} and \textit{ihfB}, we consider only one subnet,
since their regulatory action is mediated by the IHF hetero-dimer,
formed by the gene products of both genes. In contrast to older
versions of RegulonDB, release 6.3 contains eight mutual interactions
between gene pairs, and one 3-node cycle (see Materials and Methods), leading
to subnets with many shared downstream nodes (as shown in Figure
\ref{fig:examples}B for \textit{fnr-arcA}). An exemplary small subnet is
shown for the TF \textit{agaR} in Figure \ref{fig:examples}B. It
contains no regulators and can thus be depicted as a tree with only
two hierarchical levels: the root node \textit{agaR} at the top and
all ten target nodes in the bottom layer.

\subsection*{Subnet usage}
\begin{figure*}[!ht]
  \centering
 \includegraphics[width=2\columnwidth]{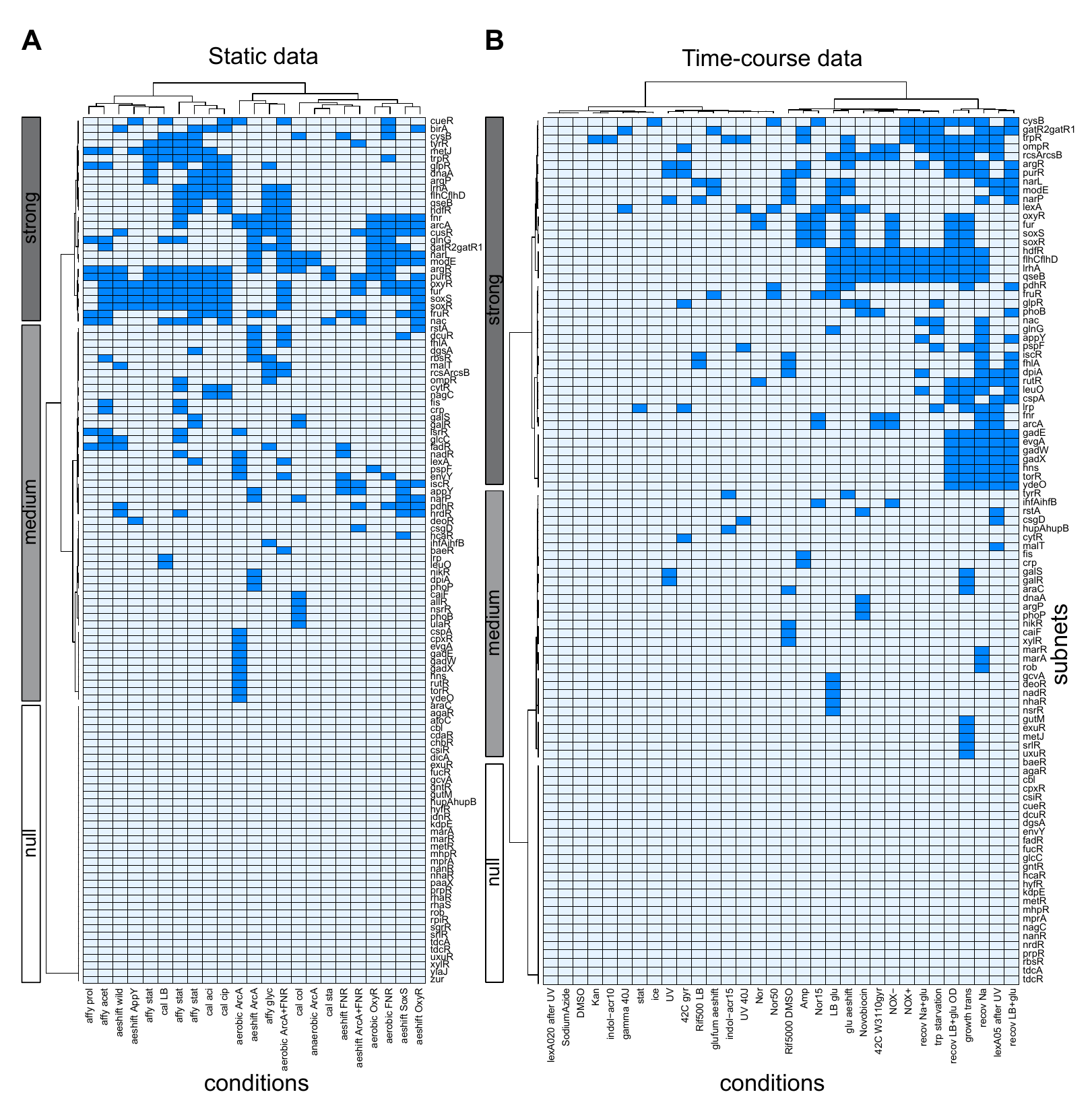}
  \caption{{\bf Subnet usage matrices.} The subnet usage matrix
    consists of subnets (rows) and conditions (columns) for the static
    ASAP data (A) and the time-course data of Sangurdekar et al. 
    \cite{sangurdekar06_genome-biol} (B). A deep blue entry
    represents a subnet significantly enriched with differentially
    expressed genes under the respective conditional change (A), or a
    subnet with collectively responding genes during the given
    time-course (B), respectively.}
  \label{fig:matrix}
\end{figure*}

We want to analyze the importance of subnets as information-processing
units in the TRN. To this end, we map large-scale expression profiles
from microarray experiments onto the TRN.  First, we consider a data
set where either wild-type \textit{E.~coli} strains are compared to
strains with genetic alterations and with cells under environmental
stress, or wild-type and mutant strains are compared under aerobic and
anaerobic growth conditions. We will refer to this data set as the
static data (see Materials and Methods for a detailed description of
the data used). For each condition, we identify differentially
expressed genes (with a statistical analysis of microarrays as introduced in
\cite{tusher01_proc-natl-acad-sci-u}, $\mathrm{FDR} \le 0.3$, see
Materials and Methods) and determine subnets significantly enriched
(Fisher's exact test at $\mathrm{FDR} \le 0.3$, see Materials and
Methods) with those genes. In Figure \ref{fig:matrix}A, we plot a
hierarchically clustered (see Materials and Methods for clustering
details) subnet usage matrix, where a deep blue entry represents a
subnet significantly enriched with differentially expressed genes.

For example, the comparison of wild-type and \textit{fnr} mutant
strains under aerobic growth conditions (denoted as `aerobic FNR' in
the usage matrix labels) yields 17 subnets with enriched
differentially expressed members: \textit{arcA}, \textit{argR},
\textit{birA}, \textit{cueR}, \textit{cusR}, \textit{cysB},
\textit{envY}, \textit{fnr}, \textit{fur}, \textit{gatR2gatR1},
\textit{glnG}, \textit{modE}, \textit{narL}, \textit{oxyR},
\textit{pdhR}, \textit{purR}, \textit{trpR}. We assume that these
subnets are directly associated with the \textit{fnr} deletion, either
due to the TF action of FNR (the roots \textit{arcA}, \textit{narL},
and \textit{pdhR} are direct targets of FNR) or via signal
transduction cascades induced by the presence or absence of
FNR. Interestingly, not all subnets embedded in the \textit{fnr}
subnet show significant differences in the expression of their
genes. These phenomena may occur due to missing data in
RegulonDB or due to interactions that rely on specific conditions and are
not active under aerobic growth (like co-activators or TF
conformations).  A functional hypothesis is that the downstream genes
of the respective root are collectively shielded from the rest of the
\textit{fnr} subnet, or that the regulatory control of the respective
node exceeds the pure promoter binding mechanism (as the analog
control \cite{marr08_bmc-syst-biol} of the known architectural protein
H-NS, see also e.g. \cite{travers05}).

The overall pattern of subnet usage for the different conditions is
rather homogeneous for all compared profiles: between 6.8\% and 20\%
of the subnets are used in each condition. However, we find a
hierarchy of usage at the subnet level and coordinately used
subnets. A clustering of subnets with respect to their subnet usage will
be discussed in the next section.

We want to compare our results with another, fundamentally
different data set consisting of time series, and an independent
analysis approach based on the collectivity of a subnet's
response. The data used in \cite{sangurdekar06_genome-biol} contains
time courses of \textit{E.~coli} transcriptome responses to diverse stimuli (like
UV and gamma radiation, norfloxacin, and different concentrations of
indol-acrylate), measured with whole-genome DNA microarrays. For each
time series we quantify the collectivity of the response of the
subnet's genes and compare it to randomly sampled subnets by
calculating the Shannon entropy of the eigenvalues of a singular value
decomposition (see Materials and Methods for details). Subnets responding
collectively are marked in Figure \ref{fig:matrix}B as deep blue
entries.

During the different time courses, the subnet usage varies between no
subnet usage at all (0\%) and a maximum of 26\%. The first 14
experiments in the matrix (including all radiation exposure
experiments and indol-acrylate treatments in different concentrations)
exhibit a subnet usage below 0.5\%. Apparently, for these experiments,
\textit{E.~coli} masters the adaptation to the imposed stress with other forms of
regulatory control. In experiments where subnets are more frequently
used, we find again blocks of collectively used subnets that differ
between sets of experiments.

\subsection*{Clustering}
Using hierarchical clustering,  we  identify clusters of subnets with distinctly
different patterns of subnet usage in both data sets.  In the subnet usage matrix
derived from the static data (Figure~\ref{fig:matrix}A), a substantial
part of the subnets are never significantly enriched with
differentially expressed genes, further on called the `null'
cluster. On the contrary, subnets in the `strong' cluster are on
average used in 25\% of the experiments. The `medium' cluster in
between has an average usage of 6.0\%.  In the time-course data
matrix (Figure~\ref{fig:matrix}B), we also identify three clusters with
markedly different subnet usage, further similarly  denoted as `strong' (20\%
average subnet usage), `medium' (3.8\%), and `null' (0.0\%).

\begin{figure}[!ht]
  \centering
  \includegraphics[width=\columnwidth]{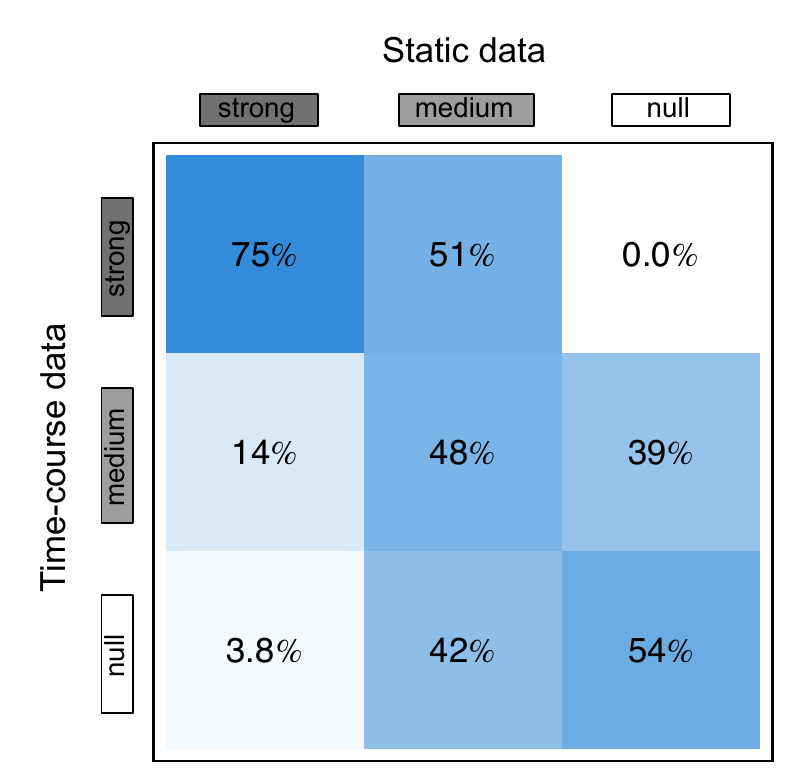}
  \caption{{\bf Subnet cluster overlap.} Relative overlap between the
    different subnet clusters of the static data and the
    time-course data. The relative overlap is calculated as the number
    of subnets present in both clusters, divided by the smaller total
    number of subnets in the two clusters under consideration.}
  \label{fig:overlap}
\end{figure}

The overlap---with respect to the subnet roots---between clusters
from the two data sets is shown in Figure \ref{fig:overlap}. We find
that the clusters in the different experiments often share subnets
(75\% for the `strong', 48\% for the `medium', and 54\% for the
two `null' cluster). Only some subnets in the `strong' cluster of the
static data appear in the `null' cluster of the time-course data (3.8\%
overlap), the static `null' cluster and the time-course `strong'
cluster are disjunct (0\% overlap). The fact that the cluster
composition differs between the two data sets may rely on the
different external stresses applied. Maybe even more importantly, in
the time-course data, an \textit{E.~coli} colony adapts spontaneously
to a environmental change applied. In contrast, strains that have
already adapted to a different environment or a genetic mutation are
compared in the static data. Still, the large overlap between the
clusters, derived from experiments with independently sampled
environmental conditions, is remarkable.

To assess the cluster composition from a functional perspective and
detect biological plausible components, we conducted a gene ontology
(GO) enrichment analysis (see Materials and Methods for details). On
the level of subnet roots, we find no enriched GO terms at all. If we
include the nodes within the subnets in each cluster, we find
several enriched categories. In the `strong' clusters of the static
and time-course data, `iron ion binding' and the `generation of precursor
metabolites and energy' appears. The less overlapping `medium' clusters
share no enriched annotations. The `null' clusters, finally, share
enriched metabolic processes (carbohydrate, fucose, D-gluconate) and
transporter activity (carbohydrate, sugar).

\begin{figure*}[!ht]
  \centering
 \includegraphics[width=2\columnwidth]{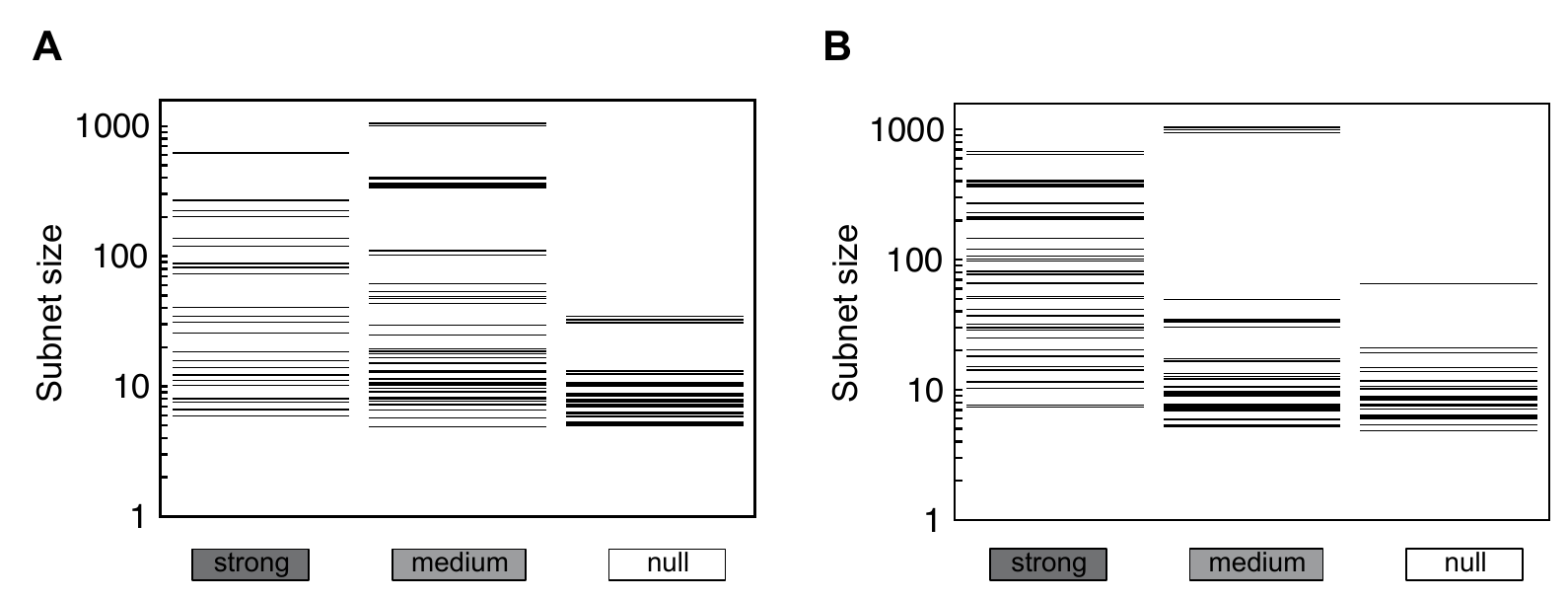}
  \caption{{\bf Subnet size composition of the clusters.} For the
    static data (A) and the time-course data (B), the `null' cluster
    is composed of subnets with less than 100 nodes, while all other
    clusters contain both small and large subnets.}
  \label{fig:sizes}
\end{figure*}

We study the sizes of the subnets contained in the different clusters
and find that the subnet composition is highly heterogeneous in both
the static (Figure \ref{fig:sizes}A) and the time-course data
(Figure \ref{fig:sizes}B): while the `strong' and `medium' clusters
contain subnets of all size, including large subnets with hundreds of
nodes spanning most of the TRN, the `null' clusters contain
preferentially small subnets with only tens of nodes (see Figure
\ref{fig:sizes}).
%rerevision
Similarly, the out-degrees of the subnet roots substantially
differ. While the master regulators \textit{fnr} in the `strong' and \textit{crp} in the
`medium' cluster control 275 and 418 nodes, respectively, the maximum
out-degree of `null' cluster subnets is 28 for \textit{marA} in the static
case, and 57 for \textit{cpxR} in the time-course data.

\subsection*{Motifs}
\begin{figure*}[t]
  \centering
  \includegraphics[width=2\columnwidth]{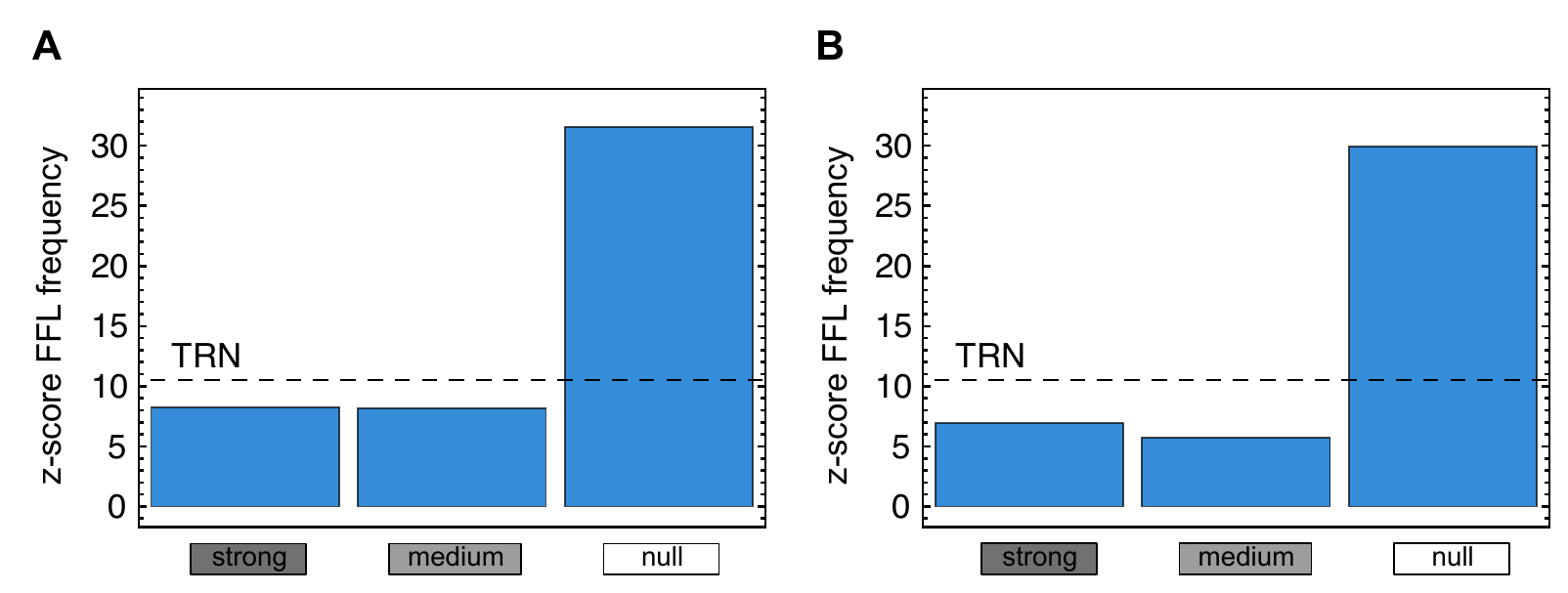}
  \caption{{\bf Feed-forward loop enrichment.} Analysis of the
    feed-forward composition of the subnet clusters identified in the
    static data (A) and the time-course data (B) respectively. In both
    data sets, we find that the z-score of the feed-forward loop
    composition is highest in the `null' cluster induced by
    non-responding subnets.}
  \label{fig:ffls}
\end{figure*}

Can we infer topological differences between the conditionally used
subnets and the unused subnets in the `null' cluster beyond subnet
size and a root's out-degree? We analyzed the 3-node motif composition
of the subgraphs induced by the subnets of each cluster (see Figure 3
in Text S1) by computing the z-score (see Materials and Methods for a
detailed description of the z-score calculation) with respect to
randomized graphs \cite{shen-orr02}. All subnets show a normalized
triad significance profile \cite{milo04} characteristic for bacterial
regulatory networks (see Figure 4 in Text S1). However, consistently
in both data sets we find in the `null' cluster an enrichment of
feed-forward loops, a well-studied motif with interesting
dynamical properties (see Figure \ref{fig:ffls}). Depending on the
actual design as a coherent or incoherent feed-forward loop, this
motif can serve as a sign-sensitive delay or an accelerator in
transcriptional networks \cite{mangan03_pnas}. Here, the feed-forward
loop z-scores of $31.6$ and $29.9$ for the static data and the
time-course data `null' cluster, respectively, distinctly exceed the
z-score of the feed-forward loop in the full TRN ($10.5$). The
z-scores of all other clusters lie below this threshold (see Figure
\ref{fig:ffls}).

To check whether enriched feed-forward loops are an artifact of the
cluster-induced subgraph sizes, we apply two null models to the static
data: First, we induce a subgraph of the TRN by randomly sampling the
same number of nodes as contained in the `null' cluster (that is,
221). Second, we randomly choose the same number of subnets as
contained in the `null' cluster (that is, 30) and therein induce
subgraphs with a size distribution similar to the one in the `null'
cluster. We generate 100 samples and find that the feed-forward loop
z-score of the `null' cluster exceeds both null model averages
($\mathrm{p} \le 0.01$ and $\mathrm{p}=0.01$, respectively, see Figure 5 in Text S1). This
indicates that the feed-forward loop enrichment is a specific property of
the identified `null' cluster and no size effect.

We test the robustness of our finding with regard to the data used in
two different ways: First, we apply a meta analysis on the 466
\textit{E. coli} experiments available in the Many microbes microarray
database \cite{faith08_nucleic-acids-res}. We analyze this data with
the entropy approach by interpreting the set of experiments as a time
series. Interestingly, the homogeneously responding subnets show no
distinct feed-forward loop enrichment ($\mathrm{z} = 8.3$) while the subnets with no
coordinated response are, similarly to the `null' cluster subnets,
highly enriched with feed-forward loops ($\mathrm{z} = 51$). Second,
to test the robustness of our findings against incomplete data, we
implement the time-course data analysis on the last four version of
RegulonDB (6.1 - 6.4). We find that irrespective of the RegulonDB
version used in our analysis, a prominent feed-forward loop enrichment in the null
cluster appears (see Figure 6 in Text S1). Notably, the number of
vertices (V) and links (L) in the TRNs increased considerably from
$V=1468, L=3040$ (RegulonDB 6.1) to $V=1540, L=3223$ (RegulonDB 6.4).

\begin{figure*}[t]
  \centering
  \includegraphics[width=2\columnwidth]{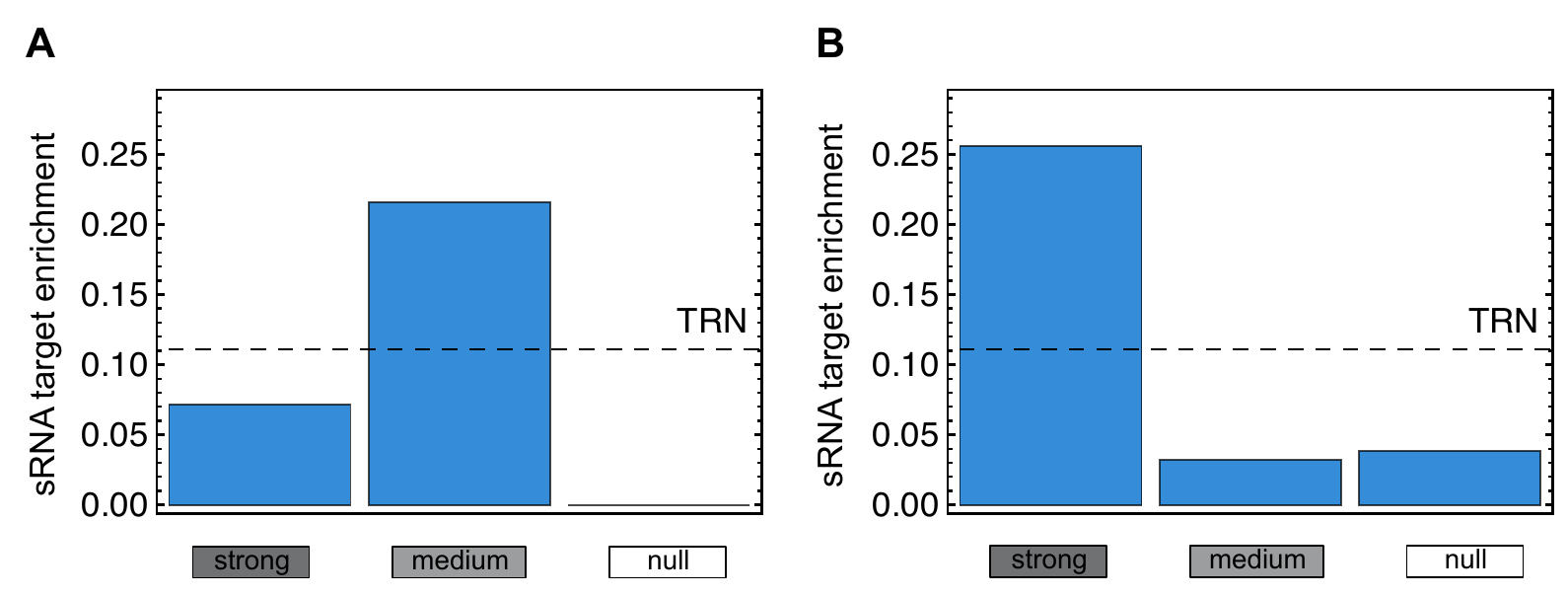}
  \caption{{\bf sRNA target enrichment.} Relative number of subnets
    with enriched sRNA targets in each cluster in the static data (A)
    and the time-course data (B) respectively. In the full TRN, we
    find 13 out of 117 subnets significantly enriched with sRNA
    targets (Fisher's exact test with $\mathrm{FDR} \le 0.05$),
    resulting in an average sRNA target enrichment of 0.11 (dashed line). In
    both data sets, we find that subnets in the `null' cluster are
    depleted with sRNA targets.}
  \label{fig:snRNA}
\end{figure*}

\subsection*{Small RNA target enrichment}
A rather recently discovered mechanism of regulatory control are small
noncoding RNAs (sRNAs) \cite{gottesman04_annu-rev-microbiol}. In
\textit{E.~coli}, up to 100 sRNAs may exist \cite{vogel05_biol-chem},
primarily as regulators of mRNA stability and translation. We first
investigate the sRNA mediated control on network motifs. Comparing the
number of 3-node motifs with at least one sRNA target with randomly
sampled sets of targets of the same size, we identify seven motifs
with significantly ($\mathrm{z} \ge 2$) enriched occurrence of sRNA
targets (see Figure 7 in Text S1). Among them, we find the
feed-forward loop (motif ID 38), and a motif (ID 110, see Figure 7 in
Text S1), which has been implicated previously with an enrichment of
microRNA targets in a mammalian signaling network
\cite{cui06_mol.-syst.-biol}.

To infer the interplay between subnet mediated control and sRNA
regulation, we map the target transcripts from RegulonDB 6.3 onto the
TRN and infer 13 subnets with a significantly enriched (Fisher's exact
test, $\mathrm{FDR} \le 0.05$, see Materials and Methods) number of sRNA target genes:
\textit{arcA}, \textit{cspA}, \textit{envY}, \textit{evgA},
\textit{fnr}, \textit{gadE}, \textit{gadW}, \textit{gadX},
\textit{hns}, \textit{ihfAihfB}, \textit{rutR}, \textit{torR},
\textit{ydeO}. In relative numbers, we find that 11\% of all TRN
subnets are enriched with sRNA targets. With regard to the clusters of
different subnet usage, enriched subnets are intriguingly absent in
the cluster with unused subnets: We find no enriched subnet in the
`null' cluster of the static data, and only one (\textit{envY}) in the
time-course data (see Figure \ref{fig:snRNA}). At the same time,
enriched subnets are present in the medium and strong cluster,
respectively. 

We draw two important conclusions from that finding:
First, the clusters inferred from the subnet usage analysis establish
categories on the set of subnets that appear to have markedly
different topological and regulatory properties. Second, regulatory
control on the subnet level coincides with sRNA mediated control,
while feed-forward loop dynamics seems less dependent of the impact of
sRNA.

\section*{Discussion}
The rationale of our analysis has been to explore the internal logic
of gene regulation by looking at different scales within the
transcriptional regulatory network of \textit{E.~coli}. The
post-transcriptional regulation mediated by sRNAs coincides with the
subnet-wide control conferred by TFs. In contrast to
this correlated regulatory control, we obtain an anti-correlated
pattern for subnet usage and the occurrence of feed-forward loops:
when the scale dominates (high subnet usage) few regulatory devices on
the smaller scale are found (low feed-forward loop
occurrence). Similarly to our previous data-driven study on the
buffering of digital and analog control \cite{marr08_bmc-syst-biol},
our results indicate a systematic interplay between distinct
regulatory mechanisms. However, in contrast to the concept of analog
and digital control, there is no evidence for a balancing between the
induction of subnets and the usage of feed-forward loops. Rather, from
the static data analysis (see Figure \ref{fig:matrix}A) we see that
upon mutation of a root node of a strongly responding subnet, other
subnets compensate for the compromised control. The reason for that
may be the difference in scales: while both analog and digital control
can operate on sets of up to hundred genes, there is a huge functional
discrepancy between the genome-wide regulation of large subnets and
the dedicated dynamical tuning of few nodes by a feed-forward loop.

Our study expands previous approaches to link topological properties
of the TRN with expression profiles. Subnets as topologically defined
units of the TRN are groups of genes that deal coordinately with
conditional or environmental changes due to shared regulatory
interactions.  The `regulon' concept, where genes are pooled if they
share a common transcription factor, is extended by taking into
account the full downstream regulation instead of only the first
hierarchical layer. `Origons' \cite{balazsi05} are the subset of
subnets with no regulatory input at the root node and have been
defined on an operon-based version of the TRN (that is, genes with the
same promoter are treated as one node). Based on the assumption that
every TF is able to sense signals in the cell, the subnet notion is a natural
generalization of the origon concept: It allows for the
identification of used subnets within larger unused subnets (which may
be origons) and, vice versa, small unused subnets within larger used
origons.

In a complementary, subsequent investigation, one could study the sRNA
target enrichment and feed-forward loop usage across the different
experimental conditions, similarly to the study of
\cite{luscombe04_nature}. This would require to distinguish the
different types of coherent/incoherent feed-forward loops
\cite{mangan06_j-mol-biol} and quantify their usage. Here we
introduced the subnet notion, verified our approach with two types
of large-scale expression data, and compared distinct scales of
regulatory control in clusters with different subnet usage.

\vfill

\begin{figure}[h]
  \centering
  \includegraphics[width=.9\columnwidth]{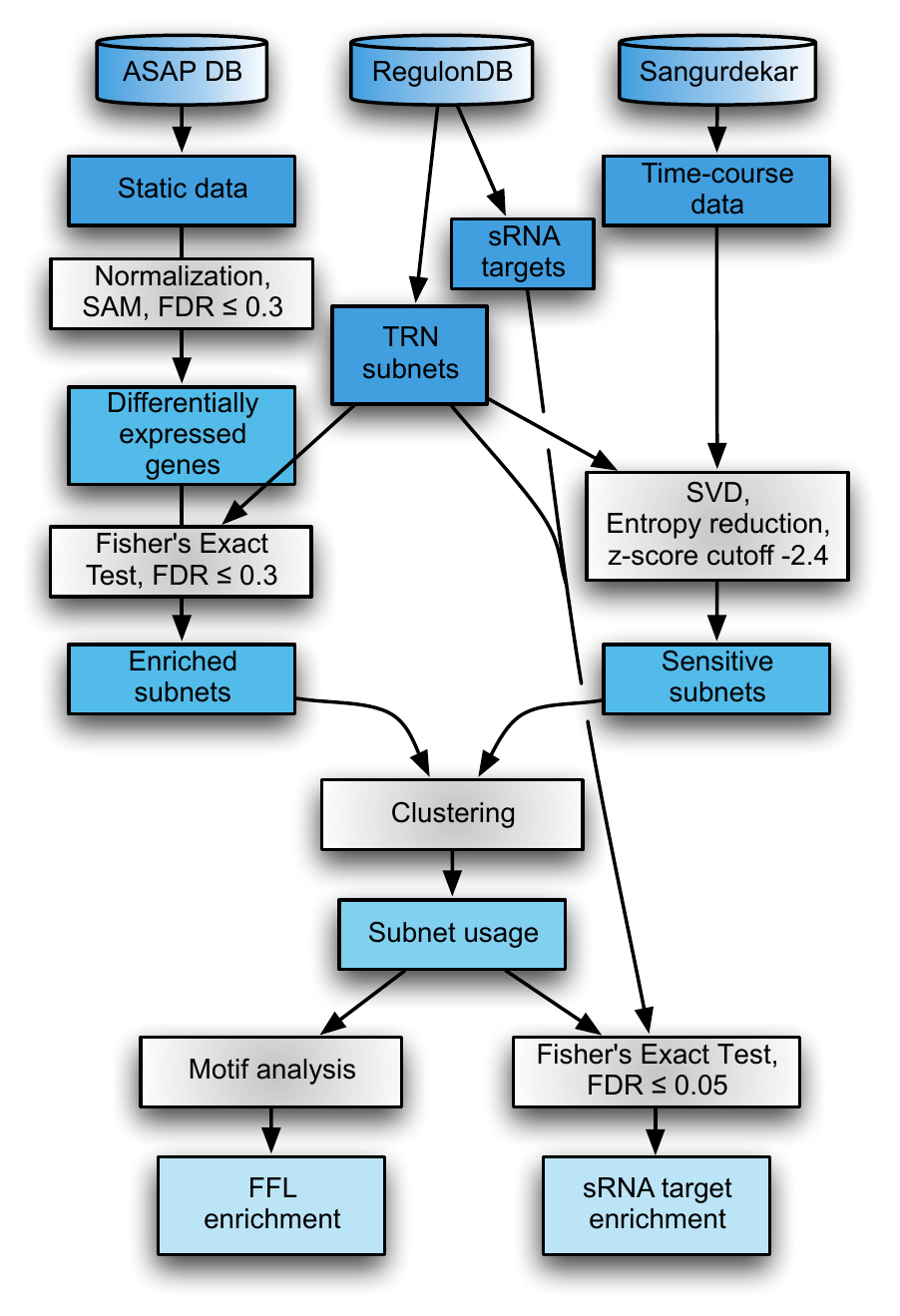}
  \caption{{\bf Workflow.} Static data and time-course data are
    analyzed differently up to the identification of
    subnets. Clustering and motif analysis is applied similarly to the resulting
    subnets.}
  \label{fig:workflow}
\end{figure}

\pagebreak

\section*{Materials and Methods}
The workflow of our analysis is illustrated in Figure
\ref{fig:workflow}, with details as follows. 

\subsection*{Network data}
We use the RegulonDB 6.3 \cite{gama-castro08_nucleic-acids-res} data
on TF-gene interactions to construct the \textit{E.~coli} TRN. Dimer
TFs (e.g.~\textit{flhCflhD} and the corresponding genes \textit{flhC}
and \textit{flhD}) are merged to a single node in the network, phantom
genes (i.e. a gene that at a previous time it was thought to be a
gene, but more recent analyses indicate it is not) are
removed. The resulting TRN comprises 1515 nodes with 3270 interactions
including 99 self-loops.

\subsection*{Subnet construction}
Within the TRN, 162 nodes have outgoing links to other nodes. The
corresponding genes confer regulatory control to other genes via TF
binding, and are called roots further on.  Each subnet is defined by a
root node and all the nodes topologically downstream. The level of a
node within the subnet is defined as the maximal distance of this node
to the root node. Pairs of genes regulating each other share the same
level. In Regulon 6.3, we find eight two node cycles ($\textit{arcA}
\leftrightarrow \textit{fnr}, \textit{crp} \leftrightarrow
\textit{fis}, \textit{gadE} \leftrightarrow \textit{gadW},
\textit{gadE} \leftrightarrow \textit{gadX}, \textit{galR}
\leftrightarrow \textit{galS}, \textit{gutM} \leftrightarrow
\textit{srlR}, \textit{marA} \leftrightarrow \textit{marR},
\textit{marA} \leftrightarrow \textit{rob}$) and one 3-node cycle
($\textit{gadW} \rightarrow \textit{gadX} \rightarrow \textit{hns}
\rightarrow \textit{gadW}$). Subnets can overlap, if they share
downstream nodes (see Figure \ref{fig:fflvssubnet}A). To allow for
significant enrichment in our expression analysis, we only consider
subnets with five nodes or more (see Figure \ref{fig:examples} for a
histogram of the subnet sizes), ending up with 117 subnets out of 162
contained within the TRN. The subnets are deposited in Text S2.

\subsection*{Expression data} 
We consider two different expression data sets to study patterns of
subnet usage.  As static expression profiles, we use Affymetrix chip
data contained in the ASAP database
({https://asap.ahabs.wisc.edu/asap/home.php}) \cite{glasner03}, namely the
data sets `Aerobic shift', `Calibrator', and `Affy data'. In each data
set, we compare different environmental (like `heat-shock') or
genotypic (like `fnr deletion') conditions with the respective
wildtype experiments, ending up with 39 chip comparisons. In the
`Aerobic shift' data set, we first calculate the estimated transcript
copy numbers (ETCNs) and compare mutant strains with wildtype strains
under both aerobic and anaerobic growth conditions.

For time-course expression profiles, we use data from
\cite{sangurdekar06_genome-biol}. There, \textit{E.~coli} strains are cultured
and subsequently analyzed on whole-genome microarrays under diverse
conditions like 'normal growth', 'suboptimal growth', 'transient
arrest', or 'severe arrest and killing'. We use 32 time-course data
sets, the number of time points varying from experiment to experiment
between 2 and 16.

\subsection*{Subnet usage}
Due to the different experimental setups we apply two
different approaches to quantify the subnet usage.

For the static data, we first determine differentially expressed
genes between two experimental conditions. For all three data sets
(`Aerobic shift', `Calibrator', `Affy data'), we compare a specific
condition with its corresponding wild-type condition (e.g., we
compare the anaerobically grown FNR deletion mutant with the
anaerobically grown wild-type strain). Additionally, we regard the
various mutants under aerobic and anaerobic conditions (e.g., we
compare the OxyR mutant expression profiles with and without oxygen
supply from the `Aerobic shift' data). In each of the 33 resulting
data set pairs (condition vs. wild-type and aerobic vs. anaerobic,
respectively) we determine differentially expressed genes by applying
the `Statistical Analysis of Microarrays' (SAM) algorithm introduced
in \cite{tusher01_proc-natl-acad-sci-u} with a Wilcoxon rank
statistics and a False Discovery Rate $\mathrm{FDR} \le 0.3$. We disregard
experiments with no genes below the significance level (10 out of
33). We then calculate a p-value for the enrichment (that is, a higher
fraction of differentially expressed genes within the subnet as
compared to the whole TRN) with Fisher's Exact Test. After multiple
testing correction, we call subnets
with $\mathrm{FDR}\le 0.3$ significantly enriched and mark them in dark blue in
the subnet usage matrix in Figure \ref{fig:matrix}A.

For the time-course data, we use a similar approach as described in
\cite{sangurdekar06_genome-biol}. To evaluate, if the genes of a
given subnet respond collectively during time to the stimulus applied,
we calculate the Shannon entropy $S$ of the normalized eigenvalues
$\varepsilon_i$ of a singular value decomposition (SVD) of the $T$
time-points vs. $L$ genes matrix, as described in \cite{alter00_pnas}:

\begin{math}
  p_i  = \varepsilon_i^2 / \sum_{k=1}^L \varepsilon_k^2\;, \,
  S = \frac{-1}{\log(L)} \sum_{i=1}^L p_i \log(p_i) \;.
\end{math}

Collectively corresponding genes give rise to a dominant principal
eigenvalue $\varepsilon$ and a small $S$.  For each subnet and each
time-course experiment, we randomly sample pseudo subnets (that is, we
randomly choose the same number of genes as contained in the
respective subnet) 1000 times and calculate a z-score (the deviation
of the subnet's entropy $S$ from the mean $\mu$ of the sampled
distribution, divided by the standard deviation of the sampled
distribution $\sigma$, $z = \frac{S-\mu}{\sigma}$). From the 117
subnets under consideration, we take only those with three or more genes
included in the data set from \cite{sangurdekar06_genome-biol},
reducing the number of analyzed subnets to 100. In order to keep the
overall number of insensitive subnets comparable to the static data
(25\%), we choose a z-score of $-2.4$ as cutoff. Collectively
corresponding subnets are marked in dark blue in the subnet usage
matrix in Figure \ref{fig:matrix}B. We validate our results with the
four latest versions of RegulonDB (6.1 - 6.4), where we keep the size
of the `null' cluster constant at 25\% and adapt the z-score cutoff
accordingly. 

As an additional meta analysis, we take the 466 \textit{E.~coli}
experiments available in the Many microbes microarray database (M3D)
\cite{faith08_nucleic-acids-res}. We analyze this data with the
entropy approach by interpreting the set of experiments as a time
series. Disregarding the principal value in the SVD, and thus
eliminating the vast chip-wide differences between the 466 experiments
included in M3D, we compare the entropy of a subnet with the entropies
of randomly sampled subnets and calculate a z-score. Similarly to our
previous analysis, we interpret subnets with z-scores below and above
the threshold $-2$ as collectively (strongly) responding and not
responding (null), respectively. From these two sets of subnets, we
induce subgraphs and calculate the feed-forward loop enrichment in the respective
graphs. Interestingly, the strongly responding subnets show no
distinct feed-forward loop enrichment ($\mathrm{z} = 8.3$) while the subnets with no
coordinated response are, similarly to the `null' cluster subnets,
highly enriched with feed-forward loops ($\mathrm{z}= 51$). Notably,
disregarding the principal value in the SVD in the analysis of the
time-course data does not alter our results. 

\subsection*{Hierarchical clustering}
From the analysis of subnet usage, we end up with two matrices, one for the
static data ($177 \times 39$) and one for the time-course data ($100 \times
32$). The matrices contain a 1 for a subnet significantly
enriched with differentially expressed genes or significantly
correlated time-courses, respectively, and a 0 otherwise. We
hierarchically cluster the two matrices using the Manhattan distance
function (see e.g. \cite{krause86_book}) and the Ward agglomerative
algorithm \cite{ward-jr63_journal-of-the-ameri}. We end up with
three clusters of subnets with clearly different usage
patterns throughout the different experiments.

\subsection*{Gene Ontology enrichment}
To infer GO term overrepresentation in the different clusters of
subnet usage, we use GOstat \cite{beissbarth04_bioinformatics} with
\textit{E. coli} UNIPROT identifiers and the `goa uniprot'
database. As parameters of the statistical test we use a p-value
cutoff of 0.01 with the Holm multiple testing correction method and a
GO-Cluster Cutoff of $-1$.

\subsection*{Motif analysis}
For each cluster, we induce a single subgraph of the
whole TRN by taking all nodes of the cluster's subnets. We thus ensure
that every motif is counted only once in each cluster. We calculate
the z-scores of the network motifs of size 3 in the TRN and the
cluster induced subgraphs with MFINDER \cite{milo02} (using 1000 random
networks). For this analysis we disregard the character of the
interaction (i.e. its activating or inhibiting impact). 

\subsection*{sRNA enrichment}
RegulonDB 6.3 contains regulatory information for 22 small RNAs and 32
target transcripts. We map these onto the TRN and find 22
target genes, within them the roots \textit{fhlA}, \textit{gadX}, and
\textit{hns}. We map the targets on the TRN subnets and calculate
the relative overrepresentation with Fisher's exact test. We correct
for multiple testing error and find 13 subnets enriched with sRNA
targets at $\mathrm{FDR} \le 0.05$: \textit{arcA},
\textit{cspA}, \textit{envY}, \textit{evgA}, \textit{fnr},
\textit{gadE}, \textit{gadW}, \textit{gadX}, \textit{hns},
\textit{ihfAihfB}, \textit{rutR}, \textit{torR}, \textit{ydeO}. For
each cluster, we calculate the relative number of subnets with sRNA
target enrichment and plot the result in Figure \ref{fig:snRNA}.

\section*{Acknowledgments}
We thank Annick Lesne (Paris, France), Nikolaus Sonnenschein (Bremen, Germany), Lina Shehadeh
(Miami, U.S.), Dominik Lutter, and Florian Bl\"ochl (both Munich,
Germany) for discussions and helpful advice. We acknowledge the comments of anonymous reviewers, that substantially
improved the manuscript.

\end{document}